
-----------------------------------------------------------------

\baselineskip=20pt\magnification=1200

\def\n{\noindent}
\def\c{\centerline}
\def\i{\item}

\def\a{\alpha}
\def\b{\beta}
\def\d{\delta}
\def\D{\Delta}
\def\dg{\dagger}

\def\ep{\epsilon}

\def\o{\omega}
\def\Om{\Omega}
\def\p{\prime}

\def\t{\tilde}
\def\th{\theta}
\line{\hfill IUCAA-4/94}
\line{\hfill January, 1994}

\vskip.3truein
\c{ SQUEEZING AND DUAL RECYCLING IN}
\c{LASER INTERFEROMETRIC GRAVITATIONAL WAVE DETECTORS}
\vskip.4truein
\c{\bf Biplab Bhawal$^*$ and Vijay Chickarmane$^\dg$}
\c{Inter University Centre for Astronomy and Astrophysics,}
\c{Post Bag 4, Ganeshkhind,}
\c{Pune-411007, INDIA.}
\vskip1truein
\c{ABSTRACT}

\n We calculate the response of an ideal Michelson interferometer
incorporating both dual recycling and squeezed light to gravitational
waves. The photon counting noise has contributions from the light
which is sent in through the input ports as well as the vacuum modes
at sideband frequencies generated by the gravitational waves.
The minimum detectable gravity wave amplitude depends on  the frequency
of the wave as well as the squeezing and recycling parameters.
Both squeezing and the broadband operation of dual recycling reduce the
photon counting noise and hence the two techniques can be used
together to make more accurate phase measurements. The variance of
photon number is found to be time-dependent, oscillating at the
gravity wave frequency but of much lower order than the constant part.

\vskip.3truein
\line{$^*$email: biplab@iucaa.ernet.in\hfill}
\line{$^\dg$email: vijay@iucaa.ernet.in\hfill}

\vfill\eject

Laser-interferometric gravitational wave detectors [1] operate by sensing the
difference in phase shifts imposed on the laser light in the two orthogonal
arms of a Michelson type of interferometer by a gravitational wave. This phase
shift manifests itself in the observed intensity change of the interference
pattern. The sensitivity of the detector is determined by two fundamental
sources of quantum mechanical noise : the photon counting error and the error
originating due to fluctuations in radiation pressure on the mirrors. At the
present level of the laser power available, the smallest detectable signal is
limited by the photon counting statistics and various efforts have been made to
increase the sensitivity level of these interferometers.

Caves[2] first realized that the photon number fluctuations at the output could
be understood due to the interference of the vacuum fluctuations of light which
enters through the unused input port of the beam splitter with the ingoing
laser light. He suggested that instead a squeezed photon state could be
injected through this port to reduce the photon counting noise. For a Michelson
interferometer operating on a dark fringe, most of the light escapes towards
the laser source. Therefore, it had been suggested [3] that this light can be
recycled by putting a mirror in front of the source to enhance the sensitivity
of the interferometer. This technique is known as Power recycling. Brillet et
al[4] argued that squeezing and power recycling are compatible with each other
and that both can be used together to improve the signal-to-noise ratio.

Gravitational waves modulate the phase of the laser light, thus generating
sidebands which, travel towards the photodetector in an interferometer
operating at the dark fringe [5]. These sidebands comprise the signal which can
also be recycled by another mirror placed in front of the photodetector. The
above method used in conjunction with power recycling is known as dual
recycling[6].

It is, therefore, important to attempt an analysis of the quantum mechanical
noise  present in a dual recycling interferometer that also uses squeezed light
and to investigate how well these two techniques work together. In this letter
we report our results obtained for interferometers operating in the broadband
mode. We arrive at a complete expression for the variance of the photon number
fluctuations which is found to have a time-dependent component. The presence of
sidebands significantly alters the noise. We calculate the minimum detectable
gravitational wave amplitude as a function of its frequency as well as
squeezing and recycling parameters and conclude that that the broad-band
operation of dual recycling is compatible with the squeezed light technique and
can therefore be used to enhance the sensitivity.

We first evaluate the minimum detectable phase difference with both squeezing
and dual recycling without considering gravitational waves. Referring to Fig.1,
monochromatic light beams (of angular frequency $\o_0$) in coherent and
squeezed vacuum states enter ports 1 and 2 respectively. The annihilation
operators $a$ and $b$ represent light in the coherent and squeezed modes
respectively. One may now write down equations for the intra-cavity electric
field operators, $E_{a^\p}$ and $E_{b^\p}$. We assume that the distance between
the recycling mirrors and the 50:50 beam-splitter has been adjusted such that
$E_{a^\p}$ and $E_{b^\p}$ add in phase with the ingoing modes, $E_a$ and $E_b$
respectively. We also assume that the beam-splitter introduces no phase shift
upon reflection for a wave incident on the side of port 2 and a phase shift
$\pi$ for a wave reflected on the side of port 1. The quantities $t_1$ ($t_2$)
and $r_1$ ($r_2$) represent the transmission and reflection coeff!
icients of the power (signal) rec

One can obtain  expressions for the annihilation operators $a^\p$ and $b^\p$ in
terms of the input modes $a$ and $b$. Then the annihilation operator describing
the mode at the output of port 2 can be given as
$${\rm Out2}:= t_2b^\p-r_2b={1\over M}[iat_1t_2\sin\th
+b\{t_2^2(\cos\th-r_1)-r_2M\}],\eqno(1)$$
where
$$M=1+r_1r_2-(r_1+r_2)=(1-r_1)(1-r_2)\eqno(2)$$
and $\th$ is the phase difference of light between the two arms of the
interferometer (at dark fringe, $\th=0$). Then the mean and the rms value of
the photon number at the output port 2 are found to be
$$N={t_1^2t_2^2\sin^2\th\over M^2}{\bar n}\eqno(3)$$
and
$$\D N=\sqrt{\bar n}{t_1t_2\sin\th\over
M^2}\bigg[t_1^2t_2^2\sin^2\th+(t_2^2\cos\th-r_1t_2^2-r_2M)^2e^{-2r}\bigg]^{1/2}
\eqno(4)$$
respectively, where $\bar n$ is the mean number of photons in the coherent beam
and $r$ is the squeeze factor. In these expressions, we have neglected terms
with coefficients $\sinh^2r$ since ${\bar n}\gg \sinh^2r$. At the dark fringe
most of the laser light escapes towards  port 1. However, for a very small
phase shift $\d\th$, we obtain a very small change in the mean number of
photons, $\d N(\th)$ at the output port 2. Equating the change to the rms
value, we, therefore obtain the minimum detectable phase $\d\th$ at a dark
fringe to be
$$\d\th={e^{-r}\over\sqrt{\bar n}}\Bigg[{t_2^2(1-r_1)-r_2M\over
2t_1t_2}\Bigg].\eqno(5)$$

As can be easily seen, for values of $r_1$ and $r_2$ close to unity and a large
squeeze factor, $\d\th$ is considerably reduced. This shows that squeezing and
recycling are compatible with each other and can be used together to increase
the sensitivity of the interferometer.

We now examine the case when a gravity wave of dimensionless amplitude,
$h(t)=h_0\sin\o_gt$, propagating along the z-axis impinges on an interferometer
whose arms are oriented along the x and y axes.
If the gravity wave interacts with the laser beam of frequency $\o_0$,
propagating along the y-axis for a time $\tau$ then the phase picked up by
light as a function of time is
$$\d\phi(t)={\o_0h_0\over 2\o_g}\int^t_{t-\tau}\sin(\o_g t)\,
dt=\ep_g\sin\o_g(t-{\tau\over 2}),\eqno(6)$$
where $\o_0$ is the laser light frequency and
$$\ep_g={\o_0h_0\over 2\o_g}\sin{\o_g\tau\over 2}.\eqno(7)$$
Due to the quadrupolar nature of the gravity wave the phase acquired by the
laser beam travelling along the x-axis is $(-\d\phi)$. The gravity wave thus
modulates the phase of light in the two arms which gives rise to a
time-dependent intensity [5].

The positive frequency part of the electric field operator propagating along
the y-axis can be written as
$$E_+(t)=\int^\infty_0{d\o\over 2\pi}\bigg({\hbar\o\over 2\ep_0A_0}\bigg)^{1/2}
a(\o)e^{-i\o t}e^{i\ep_g\sin\o_gt},\eqno(8)$$
where $\ep_0$ is the permitivity constant and $A_0$ is the cross-sectional area
of the quantization volume. One can now show[7] that after modulation, the
positive frequency part of the electric field operator can be written as :
$$E_+(t)=\int^\infty_0{d\o\over 2\pi}\bigg({\hbar\o\over 2\ep_0A_0}\bigg)^{1/2}
e^{-i\o
t}\sum^{+\infty}_{n=-\infty}\big(1+{n\o_g\over\o_0}\big)^{1/2}J_n(\ep_g)
a(\o+n\o_g),\eqno(9)$$
where $a(\o+n\o_g)$ are the annihilation operators at newly-generated
frequencies $\o\pm\o_n$ and $J_n(\ep_g)$ are the ordinary Bessel functions.

So, for any wave originally present with frequency $\o$, after modulation, one
gets sidebands at $\o\pm n\o_g$. Since $J_n(\ep_g)\sim (\ep_g)^n$ for small
$\ep_g$ and we are interested in terms of order ${\cal O}(\ep_g)$, we consider
in our calculation only the first sideband on both sides.
The sideband modes $a(\o\pm\o_g)$ and $b(\o\pm\o_g)$ which enter the
interferometer through the ports 1 and 2 respectively are all in their vacuum
states.

So, the intracavity electric field at the port 1 can be written as
$$\eqalignno{E_{a^\p}(t^\p)={e^{i\psi_1}\over 2}
&\bigg[(t_1E_a+r_1E_{a^\p}+r_2E_{b^\p}+t_2E_b)e^{-i\d\phi(t)}e^{-i\th}&\cr
&+(t_1E_a+r_1E_{a^\p}-r_2E_{b^\p}-t_2E_b)e^{i\d\phi(t)}e^{i\th}\bigg],&(10)\cr}
$$
where $\th$ is the constant phase offset between the two arms of the
interferometer; $t^\p=t+L/c$ and $L$ is twice the arm length. The phase
$\psi_1$ is that acquired by light in traversing twice the path between the
beam-splitter and the power recycling mirror.

We can now write the Fourier transform of each electric field and then pull in
the time dependent phase factor inside the integral to obtain
\vfill\eject
 $$\eqalignno{\int{\t E}_{a^\p}(\Om)e^{-i\Om t^\p}\, d\Om &=
{e^{i\psi_1}\over 2}\bigg[\bigg(
 t_1\int{\t E}_a(\Om)e^{-i\Om t}e^{-i\d\phi(t)}\, d\Om
+r_1\int{\t E}_{a^\p}(\Om)e^{-i\Om t}e^{-i\d\phi(t)}\, d\Om
&\cr &
+r_2\int{\t E}_{b^\p}(\Om)e^{-i\Om t}e^{-i\d\phi (t)}\, d\Om
+t_2\int{\t E}_b(\Om)e^{-i\Om t}e^{-i\d\phi(t)}\, d\Om\bigg)e^{-i\th}
&\cr  & +\bigg(
 t_1\int{\t E}_a(\Om)e^{-i\Om t}e^{i\d\phi (t)}\, d\Om
+r_1\int{\t E}_{a^\p}(\Om)e^{-i\Om t}e^{i\d\phi (t)}\, d\Om
&\cr &
-r_2\int{\t E}_{b^\p}(\Om)e^{-i\Om t}e^{i\d\phi (t)}\, d\Om
-t_2\int{\t E}_b(\Om)e^{-i\Om t}e^{i\d\phi(t)}\,
d\Om\bigg)e^{i\th}\bigg].&(11)\cr}$$

Now, we can easily see that the presence of $e^{-i\d\phi(t)}$ factor inside
each integral sign leads to phase modulation and subsequently to the generation
of sidebands.  We confine our attention only to the first sideband on both
sides. The constant phase factor $\th=\Om\D L$ is different for different
frequencies, but, for simplicity, we assume it to be the same for all
frequencies since the difference is very small ($\o_g\ll\o_0$). We consider the
coefficients $r_1$, $r_2$, $t_1$, $t_2$ to be independent of frequency. Since
$\o_g\ll \o_0$, we  set the factor $(1+\o_g/\o)\rightarrow 1$ in Eq.(9).

Now, one can write equations for the outgoing (primed) annihilation operators
for $\Om=\o$, $\o\pm\o_g$ after dividing throughout by the same normalization
constant.

$$\eqalign{
          a^\p(\Om)\exp[-i(\Om L/c+\psi_1(\Om))]=& Ae^{-i\th}+Be^{+i\th},\cr
          b^\p(\Om)\exp[-i(\Om L/c+\psi_2(\Om))]=& Ae^{-i\th}-Be^{+i\th},\cr
          }\eqno(12)$$
where
$$\eqalign{
           A=&  t_1\sum_n a(\Om+n\o_g)J_n(-h)
               +r_1\sum_n a^\p(\Om+n\o_g)J_n(-h)
         \cr &
               +r_2\sum_n b^\p(\Om+n\o_g)J_n(-h)
               +t_2\sum_n b(\Om+n\o_g)J_n(-h),\cr
           B=&  t_1\sum_n a(\Om+n\o_g)J_n(+h)
               +r_1\sum_n a^\p(\Om+n\o_g)J_n(+h)
         \cr & -r_2\sum_n b^\p(\Om+n\o_g)J_n(+h)
               -t_2\sum_n b(\Om+n\o_g)J_n(+h),\cr
          }\eqno(13)$$
where the index $n$ can take values 0 and $\pm 1$.
The important quantity in the above equation is the phase factor $(\Om
L/c+\psi_i(\Om))$ that appears on the left hand side. The phase $\psi_i$ will
be different for different frequencies $\Om$. However, the point to be noted
here is that we adjust the distance between the recycling mirrors and
beamsplitter in such a way that these phase factors become unity -- a condition
called `on resonance'. This esssentially means that the laser light as well as
the sidebands are resonant with the cavities formed by the recycling mirrors.
This is termed as the broad-band operation of dual recycling.

We now have six coupled equations for $a^\p$ and $b^\p$ at three different
frequencies (i.e. $\o$ and $\o\pm\o_g$). One can arrange these equations in the
following matrix form
$$P_{ki}A^\p_i=Q_{kj}A_j, \eqno(14)$$
where $P_{ki}$ and $Q_{kj}$ are two $6\times 6$ matrices and
$$\eqalign{
A^\p_i\equiv &{\rm Transpose}(a^\p_0,\, a^\p_-,\, a^\p_+,\, b^\p_0,\, b^\p_-,\,
b^\p_+),\cr
A_j\equiv &{\rm Transpose}(a_0,\, a_-,\, a_+, b_0,\, b_-,\,
b_+\,),\cr}\eqno(15)$$
where (and from now onwards) indices $0$, $(-)$ and $(+)$ correspond to $n=0$,
$-1$ and $+1$ respectively. So, for example, $a_0\equiv a(\o_0)$, $a_-\equiv
a(\o_0-\o_g)$, $b_+\equiv b(\o_0+\o_g)$ etc.

So, all the six equations can be solved and the six primed annihilation
operators can be written in terms of the six input (unprimed) annihilation
operators through a $6\times 6$ matrix, $P^{-1}_{ki}Q_{kj}$. If we define
$a_\mu$ and $b_\mu$ as two $3\times 1$ column vectors, i.e., $a_\mu={\rm
Transpose}(a_0,\, a_-,\, a_+)$ and $a_\mu={\rm Transpose}(a_0,\, a_-,\, a_+)$,
one can write the output fields at port 2 in a simple form
$$\eqalignno{c_\a=& t_2b^\p_\a-r_2b_\a&\cr
                 =& X_{\a\mu}a_\mu+Y_{\a\nu}b_\nu ,& (16)}$$
where ${ X}_{\a\mu}$ and ${Y}_{\a\nu}$ are two $3\times 3$ matrices
and (from now onwards) the greek indices take values $0$, $(-)$ and $(+)$.
The values of different components of the $3\times 3$ matrices $X_{\a\mu}$ and
$Y_{\a\nu}$ are given below
$$X_{00}=X_{--}=X_{++}=-i{t_1t_2\sin\th\over (1-r_1)(1-r_2)},$$
$$Y_{00}=Y_{--}=Y_{++}=1,\eqno(17)$$
$$X_{+-}=X_{-+}=Y_{+-}=Y_{-+}=0.$$
The components of order ${\cal O}(\ep_g)$ are
$$\eqalign{X_{+0}=&-X_{0+}=X_{0-}=-X_{-0}=\ep_g{t_1t_2\over (1-r_1)(1-r_2)},\cr
Y_{0+}=&-Y_{+0}=Y_{-0}=-Y_{0-}=i\ep_g{t_2^2(1+r_1)\sin\th\over
(1-r_1)(1-r_2)}\cr}.\eqno(18)$$

We essentially follow references [7,8] for the expression for the
time-dependent photocurrent, ${\hat N}$.
$$\eqalignno{{\hat N}=&\sum_{\mu,\nu}c^\dg_\mu c_\nu &\cr
=& X^*_{\mu\a}X_{\nu\b}a^\dg_\a a_\b+Y^*_{\mu\a}Y_{\nu\b}b^\dg_\a b_\b
+X^*_{\mu\a}Y_{\nu\b}a^\dg_\a b_\b +Y^*_{\mu\a}X_{\nu\b}b^\dg_\a a_\b.
&(19)}$$
The mean number of photons, ${\bar N}$ is made up of a constant part, ${\bar
N}_0=\sum_\a c_\a^\dg c_\a$ and the time-dependent part $\d {\bar
N}(t)=\sum_{\a\not=\beta}c^\dg_\a c_\beta$.
The latter is essentially due to the beating of modes of two different
frequencies which gives rise to the time-dependent part at $\o_g$. There is
also a time-dependent part at $2\o_g$ but of order ${\cal O}(\ep_g^2)$ and so
we neglect it. Hence we get
$$<{\hat N}>={\bar n}_0+\d I(t)={\bar
N}{t_1^2t_2^2\sin\th\over(1-r_1)^2(1-r_2)^2}[\sin\th+4\ep_g\sin\o_gt].
\eqno(20)$$
The variance in photon number is given by
$$\eqalignno{(\D {\bar N})^2=&<{\hat N}^2>-<{\hat N}>^2&\cr
&={\bar
n}\bigg[{3t_1^4t_2^4\sin^4\th\over(1-r_1)^4(1-r_2)^4}+{2t_1^2t_2^2\sin^2\th
\over (1-r_1)^2(1-r_2)^2}+{t_1^2t_2^2\sin^2\th\over (1-r_1)^2(1-r_2)^2}e^{-2r}
\bigg]&\cr
&+{\bar n}\ep_g\sin\o_gt\bigg[{20t_1^4t_2^4\sin^3\th\over(1-r_1)^4(1-r_2)^4}
+{8t_1^2t_2^2\sin\th\over (1-r_1)^2(1-r_2)^2}
-{4t_1^2t_2^4(1+r_1)\sin^3\th\over(1-r_1)^3(1-r_2)^3}&\cr &
+4e^{-2r}\Bigg({t_1^2t_2^2\sin\th\over (1-r_1)^2(1-r_2)^2}
-{t_1^2t_2^2(1+r_1)\sin^3\th\over(1-r_1)^3(1-r_2)^3}\Bigg)\bigg].&(21)\cr}$$
The terms appearing within the brackets of the constant part of Eq.(21) can be
explained as follows: the first term is due to the coherent excitations ($a_0$)
superposed with coherent fluctuations as well as with the vacuum fluctuations
of $a_{-}$ and $a_+$. The second term is coherent excitations superposed with
vacuum fluctuations from $b_-$ and $b_+$, whereas the third term is due to the
interference between squeezed and coherent light. The time-dependence of
$\D{\bar N}$ arises due to the beating of the time-dependent part of ${\bar N}$
with its constant part. Everywhere, we have neglected terms representing
squeezed fluctuations being superposed on all the vacuum fluctuations since
${\bar n}\gg\sinh^2r$. The variance being time-dependent would mean that the
frequencies separated by $\o_g$ are correlated although the spectrum is white.
This has been referred to in the literature [9] as modulated shot noise.

The minimum detectable gravity wave amplitude $h_0$ is now obtained by setting
equal the maximum value of the `signal', $\d {\bar N}(t)$ to the maximum value
of $\D {\bar N}$. This gives us a quadratic equation in $h_0$
$$\eqalignno{h_0^2=&{4\o_g^2\over{\bar n}\o_0^2\sin^2(\o_g\tau/2)}\bigg[{3\over
16}\sin^2\th+{(1-r_1)^2(1-r_2)^2\over 8t_1^2t_2^2}+{(1-r_1)^2(1-r_2)^2\over
t_1^2t_2^2}{e^{-2r}\over 16}\bigg]&\cr &
+{h_o\over {\bar n}}{2\o_g\over\o_0\sin(\o_g\tau/2)}\bigg[{5\over
4}\sin\th+{1\over 2}{(1-r_1)^2(1-r_2)^2\over
t_1^2t_2^2\sin\th}-{(1-r_2)\sin\th\over 4}& \cr &
+e^{-2r}\bigg({(1-r_1)^2(1-r_2)^2\over 4t_1^2t_2^2}
-{(1-r_1)\sin\th\over 4t_2^2}\bigg)\bigg].&(22) \cr}$$
Since $h_0$ is already very small, we neglect the term $h_0/{\bar n}$. We
finally arrive at the expression for $h_0$
$$h_0={2\o_g\over \o_0\sqrt{\bar n}\sin(\o_g\tau/2)}\sqrt{{3\over
16}\sin^2\th+{(1-r_1)^2(1-r_2)^2\over 8t_1^2t_2^2}+{(1-r_1)^2(1-r_2)^2\over
t_1^2t_2^2}{e^{-2r}\over 16}}.\eqno(23)$$
At the dark fringe ($\th=0$) the first term is negligibly small. For a large
squeeze factor $r$ and values of reflection coefficients $r_1$ and $r_2$ close
to unity, $h_0$ can be considerably reduced.

If losses are introduced in the recycling mirrors, it would be possible to
optimize $h_0$ in terms of the parameters $r$, $r_1$, $r_2$. Experimentalists
usually implement internal [10], external [11] phase modulation (at MHz
frequency) so that measurements of intensity can be made at sufficiently high
frequency where the noise is really shot noise limited. Using the equations
described above, it should be possible to include both internal (or external)
and gravity wave modulation (including losses) and calculate the minimum
detectable gravity wave amplitude. Work in this direction  is currently being
pursued and will be communicated in future[12].

\vskip.4truein

We are grateful to B.S. Sathyaprakash for his advice and many enlightening
discussions. We also thank S.V. Dhurandhar and Kanti Jotania for suggestions.

\vfill\eject
\n{\bf References}

\i{1.} R.E. Vogt et al, Caltech LIGO proposal, 1989; A. Giazotto et al, The
VIRGO project (INFN, 1989); K.S. Thorne, in {\sl 300 Years of Gravitation},
Eds. S.W. Hawking and W. Israel, (Cambridge Univ. Press, 1987).
\i{2.} C.M. Caves, Phys. Rev., D23, 1693 (1981).
\i{3.} R.W.P. Drever, {\it in} {\sl Gravitational Radiation}, Eds: N. Deruelle
and T. Piran [North-Holland, Amsterdam, 1983].
\i{4.} A. Brillet, J. Gea-Banacloche, G. Leuchs, C.N. Man, J.Y. Vinet, {\it in}
{\sl The Detection of Gravity Waves}, Ed: D.G. Blair, [Cambridge Univ. Press,
1991].
\i{5.} A. Giazotto, {\sl Phys. Rep.}, {\bf 182}, 365 (1989).
\i{6.} B.J. Meers, {\sl Phys. Rev.}, {\bf D38}, 2317 (1988); B.J. Meers, Phys.
Lett., {\bf A142}, 465 (1989); K.A. Strain and B.J. Meers, {\sl Phys. Rev.
Lett.}, {\bf 66}, 1391 (1991).
\i{7.} J. Gea Banacloche, G. Leuchs., {\sl J. Mod. Opt.}, {\bf 34}, 793 (1987).
\i{8.} H.P. Yuen, J.H. Shapiro, {\sl IEEE Trans. Information Theory}, {\bf 26},
78 (1980); R.S. Bondurant, {\sl Phys. Rev.}, {\bf A32}, 2797 (1986).
\i{9.} T.M. Niebauer, R. Schilling, K. Danzmann, A. R\"udiger, W. Winkler, {\sl
Phys. Rev. A}, {\bf 43}, 5022 (1991).
\i{10.} C.N. Man, D. Shoemaker, M. Pham Tu, D. Dewey, {\sl Phys. Lett.}, {\bf
A148}, 8 (1990)
\i{11.} R. Weiss, {\sl Progress Report, Research Laboratory of Electronics,
MIT}, {\bf 105}, 54 (1972).
\i{12.} B. Bhawal and V. Chickarmane, {\it in preparation}

\vskip.4truein
\n{\bf Figure Caption}

\n Fig.1: A schematic diagram for dual recycling. BS- Beam-Splitter, EM- End
Mirror, PRM- Power Recycling Mirror, SRM- Signal Recycling Mirror.

\vfill\eject
\end